\documentstyle[aps,preprint]{revtex}
\begin{document}
\title{Self-similarity and novel sample-length-dependence of conductance in
quasiperiodic lateral magnetic superlattices}

\author{Z. Y. Zeng$^{1,2}$ and F. Claro$^1$}

\address{1. Facultad de F\'isica, Pontificia Universidad de
Cat\'olica de Chile, Cadilla 306, Santiago 22, Chile \\
2.Department of Physics, Human Normal University, Changsha 410081,
P. R. China and CCAST(World Laboratory), P. O. Box 8730, Beijing
100080, P. R. China}

\maketitle

\begin{abstract}

We study the transport of electrons in a Fibonacci magnetic
superlattice produced on a two-dimensional electron gas modulated
by parallel magnetic field stripes arranged in a Fibonacci
sequence. Both the transmission coefficient and conductance
exhibit self-similarity and the six-circle property. The presence
of extended states yields a finite conductivity at infinite
length, that may be detected as an abrupt change in the
conductance as the Fermi energy is varied, much as a
metal-insulator transition. This is a unique feature of transport
in this new kind of structure, arising from its inherent
two-dimensional nature.

\end{abstract}



\section{Introduction}

The discovery of quasicrystals in 1984 \cite{Shechtman} has
attracted a great amount of experimental and theoretical attention
to quasiperiodic systems\cite{Kohmoto}.  It has been shown that a
1D quasiperiodic array of electric barriers is characterized by
its self-similar energy spectrum and critical (neither extended
nor localized) states. Recent advances in semiconductor and
nano-technologies have permitted the realization of a uniform
magnetic field at nanometer scales, by creating magnetic dots or
depositing ferromagnetic or superconducting material patterns on
heterostructures \cite{McCord}. The energy spectrum and transport
properties of a two-dimensional electron gas (2DEG) modulated by a
regular array of nanoscale magnetic field inhomogeneities have
been investigated both theoretically and experimentally
\cite{Peeters}.

In this paper we discuss electron motion in a 2DEG subject to the
field of a nearby quasiperiodic array of parallel magnetic
stripes. This case differs from electric or dielectric modulation
in that
 2DEG electron tunneling through magnetic barriers is inherently a
two-dimensional process \cite{Peeters}. The effective potential
experienced by the electron is dependent on the wave vector
perpendicular to the tunneling direction. As we show below, since
for a quasiperiodic magnetic pattern this potential is still
quasiperiodic for any given transverse wavevector, both the
transmission and conductance display quasiperiodic properties. Our
main finding, however, is that the presence of extended states
somewhere in the spectrum produces a residual conductivity at
infinite length, which is lost as the incident energy decreases,
much in the manner of a metal-insulator transition.

\section{Formulaton}

We consider a $2DEG$ under an inhomogeneous perpendicular magnetic
field produced by two types of magnetic blocks $P$ and $Q$
arranged in a Fibonacci sequence (Fig. 1 (a)). The magnetic field
is assumed to be uniform along the $y$ direction and to vary along
the $x$ direction. Throughout this work we use the Landau gauge
${\bf A}=(0,A(x),0)$. For magnetic block $P/Q$ of width
$L_{P/Q}=d_{P/Q}+l_{P/Q}$, we assume for simplicity a magnetic
profile $B(x)=B_{P/Q}l_B[\delta(x)-\delta(x-d_{P/Q})]$ (Fig. 1
(b)). Its corresponding vector potential can be chosen as
$A(x)=B_{P/ Q}l_B[\theta(x)-\theta(x-d_{P/Q})]$ (Fig. 1 (c)),
where $l_B=\sqrt{\hbar /eB_P}$ and $\theta(x)$ is the Heaviside
step function.  By introducing in addition the cyclotron frequency
$\omega _c=eB_P/m^{*}$ ($m^*$ is the effective mass of electrons),
all quantities below are transformed into dimensionless units. For
$GaAs$ and an estimated $B_P=0.1T$, then $l_B=81.3 nm$,
$\hbar\omega_c=0.17 mev$ and $l_B\omega_c=1.4 m/sec$. Writing the
wavefunction in the form $e^{iqy}\psi (x)$ ($q$ is the wavenumber
associated with the spatial degree of freedom in the direction of
the stripes), one obtains the following $1D$ Schr\"{o}dinger
equation governing the motion of $2DEG$ electrons in the presence
of the magnetic modulation,
\begin{equation}\label{schrondinger}
\{\frac{d^2}{dx^2}-f_j(x)[f_j(x)+2q]\} \psi (x)=2(q^2/2-E)\psi(x).
\end{equation}
Here $f_j(x)$ is an oscillating function arising from the
Fibonacci sequence $S_j$ constructed from the vector potentials
$A_P$ and $A_Q$, and $V(x,q)=f_j(x)[f_j(x)+2q]/2$ can be
considered as an effective q-dependent potential for motion along
the tunneling direction. The dependence on $q$ of the quantity
$V(x,q)$ implies that this problem is inherently two dimensional.
Here and in what follows, we assume $B_P \ge B_Q$. Then, in the
units chosen, the function $f_j(x)$ is a sequence of barriers of
height $r=B_Q/B_P \leq 1$. For a given $q$, electron tunneling
through the magnetic structure will be equivalent to electron
motion in a $1D$ Fibonacci electric potential with square barriers
($q \ge -r/2$), square barriers and wells ($-1/2< q<-r/2$), or
square wells ($q \le -1/2$).

Matching wave functions at the edges of the magnetic block $Q$
yields the following transfer matrix $M_Q$ for an electron
propagating through such block,
\begin{equation}\label{transferp}
M_Q=\biggl(\matrix{ [\cos (k_Q d_Q)-i \mu^+_Q \sin (k_Q d_Q)]e^{-i
k_0 l_Q}& -i \mu^-_Q \sin(k_Q d_Q)e^{ik_0 l_Q} \cr \cr i \mu^-_Q
\sin(k_Q d_Q)e^{-ik_0 l_Q} & [\cos(k_Q d_Q)+i\mu^+_Q \sin(k_Q
d_Q)]e^{ik_0 l_Q} \cr }\biggr),
\end{equation}
where $k_Q=\sqrt{2E-q^2-r(r+2q)}$, $k_0=\sqrt{2E-q^2}$, and
$\mu^{\pm}_Q=\frac12(k_Q/k_0 \pm k_0/k_Q)$,  The transfer matrix
$M_P$ can be obtained by the replacements $Q \rightarrow P$ and
$r\rightarrow 1$ in the above Eq.(\ref{transferp}).

A Fibonacci multilayer system $S_j$ has $F_j$ layers, where $F_j$
is a Fibonacci number satisfying the recursion relation
$F_{j+1}=F_j+F_{j-1}$ $(j\geq 1)$, with $F_0=F_1=1$. Then
$M_{j+1}=M_jM_{j-1}$ $(j\le 1)$, with initial condition $M_0=M_Q$,
$M_1=M_P$, which yields a trace map $x_{j+1}=2x_jx_{j-1}-x_{j-2}$
and a constant of motion
$I=x^2_{j+1}+x^2_j+x^2_{j-1}-2x_{j+1}x_jx_{j-1}-1$ \cite{Kohmoto},
where $x_j=TrM_j/2$. The constant of motion $I$ characterizes the
extent of quasiperiodicity of the Fibonacci system.

Now we consider a simple case, i.e., $r=1$, $d_P=d_Q=d$, which is
likely to be the easiest to realize experimentally. Then one has
$k_P=k_Q=k$ and the constant of motion
\begin{equation}\label{motion}
I=\{(k^2-k_0^2)\sin(kd)\sin[k_0(l_Q-l_P)]/2kk_0\}^2.
\end{equation}
For the case $l_P=l_Q$, $I=0$, which corresponds to a purely periodic
magnetic superlattice. According to Eq.
(3) $I$ is also dependent on the normal wavevector $q$ through
$k_0$ and $k$. It is in general positive definite for most
incident energies $E$ if $l_P \ne l_Q$. One thus expects the
quasiperiodic self-similarity to appear in the energy spectra,
transmission and, possibly, the conductance. In terms of the
matrix $M_j$  the transmission coefficient becomes
\begin{equation}
T(E,q)=4/[Tr(M_j^t M_j)+2],
\end{equation}
where the superscript $t$ denotes the transpose of a matrix. The
conductance $g$ is calculated from the Landauer-B\"uttiker formula
\cite{Buttiker} by averaging the electron flow over half the Fermi
surface,
\begin{equation}
g=\int_{-\pi/2}^{\pi/2}T(E,\sqrt{2E}\sin\theta)\cos{\theta}
      d\theta.
\end{equation}
Here  $\theta$ is the angle between the velocity of incidence and
the tunneling axis $x$, $E$ is the incident energy, and the
conductance is expressed in units of the quantity
$e^2m^*vl_y/\hbar^2$, where $v$ is the velocity of the incident
electrons and $l_y$ the width of the sample.

\section{Results and discussion}

We  show in figure 2 typical transmission spectra for different
transverse wavenumbers  and Fibonacci sequences $S_9, S_{12}$ and
$S_{15}$. The magnetic structure parameters are chosen as $r=1,
d_P=d_Q=1, l_P=1, l_Q=2$. From the left to the right column, $q$
is $-0.7$, $0.0$ and $0.7$, which corresponds to the equivalent
Fibonacci electric superlattices constructed from two square-well,
two low square-barrier and two high square-barrier blocks,
respectively. From the top to bottom row the transmission spectra
are for $S_9, S_{12}$ and $S_{15}$. respectively. As clearly shown
in figure 2, the transmission spectra for $S_9, S_{12}$ and
$S_{15}$ are self-similar, i.e., the transmission peak clusters
and the transmission gaps for different Fibonacci sequences are
arranged in a very similar way , regardless of the value of the
transverse wave-number $q$. In fact, the self-similar transmission
spectra are the refection of the  self-similarity in the
corresponding energy spectra (not shown).

Figure 3 illustrates the self-similarity of the transmission
spectra more clearly. The first and second rows show that,
regardless of the value of $q$, the transmission bands are
tribranching hierarchically in a self-similar way. It is the
self-similarity between the whole and the local spectra. Also one
can readily observe the similarity between the transmission
spectra of $S_{12}$ and $S_{15}$ at quite different scales. The
third column shows in more detail this scaling property at $q=0$
as the length is increased. We notice in this data that the
evolving structure has a six-circle symmetry, arising from the
property $M_{j+6}=M_j$. In fact, the scale change of the incident
energy $E$ between the spectra for $S_9$, $S_{12}$ and $S_{15}$ is
given by the scaling index of the renormalization group
transformation of the six-circle map $[1+4(1+I)^2]^{1/2}+2(1+I)$
\cite {Kohmoto}. The self-similarities of the transmission spectra
arise from the self-similar energy spectra (not shown) of this
special structure. Close inspection of the data in the third
column shows in addition that there are states with transmission
coefficient equal to unity, that persist as the length is
increased (arrows in the figure) \cite {Azbel,Sire}. These exotic
extended states play a crucial role in the unusual length
dependence of the conductance described below.

Numerical results for the conductance for structures $S_{9}$,
$S_{12}$ and $S_{15}$ are shown in Fig. 4.  The first column shows
the development of further fine structure each time the Fibonacci
number increases, keeping the position of the main dips at each
step roughly unchanged. Although according to (5) the conductance
is an average of transmission coefficients over half the Fermi
surface, self-similarity is still present, as made evident in the
next two columns. In the center column we repeat the central panel
of the first column in order to show how a change of scale
produces a similar spectra in $S_{12}$, while an increase in
length ($S_{15}$) gives more detailed structure, also similar to
the whole pattern. The column in the right illustrates the
six-circle scaling property of the conductance.

Close inspection of Fig. 4 also shows that as the length of the
sample increases, the conductance decreases. To find out what
governs this behavior we have calculated the conductance at
incident energies $E=0.125, 0.30, 0.45$ and $0.60$ for different
sample lengths $l_x$, and plotted them in the upper ($r=0.5$) and
bottom panels ($r=1.0$) of Fig. 5. We find that in all cases the
conductance dependence in $l_x$ is bounded from below by a power
law decrease \cite{Sutherland}. In order to investigate the
possibility of a residual conductivity at infinite length, we
include in the figures a fit using the function $g = g_0 e^{\beta
l_x^{-\alpha}}\sim g_0(1 + \beta l_x^{-\alpha})$. We notice that
in the upper panels the residual conductivity $g_0$ rises abruptly
by four orders of magnitude when increasing the energy from $.30$
to $.45$. Closer study of this range shows that the rise occurs
between $.43$ and $.45$. We interpret the change as the capturing
of one or more conducting channels by the convolution (5), arising
possibly from exotic extended states. The bottom panels, showing
the special case $r=1$, permit a check of this ansatz. Since the
conductance is a convolution over a range of wavenumbers $q$, if
at a particular value of this quantity an extended state is
present, it will contribute to transport at infinite length. As
may be easily checked, when $r=1$ the effective potential in Eq. 2
vanishes everywhere at wavenumber $q_c=-1/2$, a state captured by
the convolution (2) at energies $E>0.125$. The bottom panels show
that a drop by several orders of magnitude occurs when this energy
is approached from above, confirming that the loss of an extended
state is indeed reflected in a large change in the conductance.

This novel behavior is different from either the usual $1D$
\cite{Sutherland} or the $2D$ \cite{Ueda} Fibonaci systems. We
attribute it as a manifestation of the presence of exotic extended
states within the spectrum of a Fibonacci structure. Since the
conductance is a convolution over a range of wavenumbers $q$, if
at a particular value of this quantity an extended state is
present, it will contribute to transport at infinite length. When
such conducting channels are present, one may write $g\sim
g_{0}+g_{c}$, exhibiting the contributions from the critical and
the extended states. Since $g_0$ has no dependence on $l_x$ and
$g_c \propto g_{\alpha}l_x^{-\alpha}$ \cite{Ueda}, the conductance
$g$ may be approximated by the function
 $g_0 e^{\beta l_x^{-\alpha}}$,
 where $\beta \propto g_{\alpha}/g_0$. It can be expected $\beta > 1$
 due to the predominant weight of the critical channels.

The self-similarities and the length-dependence of the
transmission and conductance of a Fibonacci magnetic superlattice
reported above are robust with regard to changes in the particular
shape of the magnetic barriers, and the choice of vector potential
\cite{unpublished}. This makes an experimental verification of the
properties found very plausible. As our results suggest, a 2DEG
subject to the inhomogeneous magnetic field of a Fibonacci or
other quasiperiodic sequence of  magnetic stripes deposited on a
nearby parallel surface should exhibit self-similarity and an
unusual length-dependence in the conductance perpendicular to the
stripes. Extended states along the direction perpendicular to the
stripes may contribute or not to the bulk conductance depending on
the energy of the incoming electrons. A possible experimental test
of this finding is to measure the conductance as the Fermi energy
is varied by means of a gate voltage. The loss of extended states
within the energy range available for transport would reveal a
drop akin to a metal-insulator transition.

\section{conclusion}

In summary, we have discussed the quasiperiodic behavior of
electrons in a Fibonaci lateral magnetic superlattice. We have
shown that its transmission and conductance possess both the
self-similarity and six-circle properties found in other kinds of
quasiperiodic systems. Moreover, novel scaling properties of
conductance with respect to the sample size in the tunneling
direction have been found, exhibiting the presence of exotic
extended states.

\begin{acknowledgments}
This work was supported in part by a C\'atedra Presidencial en
Ciencias and FONDECYT 1990425 (Chile), and  NSF grant No.
53112-0810 of Hunan Normal University (China). We are indebted to
J. Bellissard for useful comments and suggestions relating to our
results. Discussions with W. Yan and L. D. Zhang and
communications with F.M. Peeters and M. B\"{u}ttiker are also
acknowledged.

\end{acknowledgments}


\begin{figure}

\caption { The Fibonacci magnetic superlattice (a), showing the
magnetic profile $B(x)$ of building blocks $P$ and $Q$ (b), and
the corresponding y-component of the vector potential, A(x)(c).}
\end{figure}

\begin{figure}
\caption {Transmission spectra of Fibonacci magnetic superlattices
$S_{9}$, $S_{12}$, $S_{15}$ (from top to bottom)  for $q=-0.7$
(left column), $q=0.0$ (middle column) and $q=0.7$ (right column).
The magnetic structure parameters are $r=1, d_P=d_Q=1, l_P=1,
l_Q=2$.}
\end{figure}

\begin{figure}
\caption {Transmission spectra of Fibonacci magnetic superlattices
$S_{12}$, $S_{15}$  for $q=-0.7$ (left column), $q=0.7$ (middle
column) and $S_{9}$, $S_{12}$, $S_{15}$ for $q=0.0$ (right
column). The magnetic structure parameters is the same as in Fig.
2}
\end{figure}

\begin{figure}
\caption{Conductance of Fibonacci magnetic superlattices $S_{9},
S_{12}, S_{15}$. The magnetic structure parameters are the same as
in Fig. $2$, except $B_P=B_Q=2$ for the right column, set to
discern the subtle structure.}
\end {figure}

\begin{figure}
\caption{ Length dependence of the conductance. $B_P=2B_Q$ (upper
panels) and $B_P=B_Q$ (bottom panels), and the other parameters
are the same as in Fig. $2$. The solid lines are an exponential
fit described in the text.}
\end {figure}

\end{document}